\documentclass[12pt]{article}
\def\tr{\hbox{Tr}}

\def\be{\begin{eqnarray}}
\def\ee{\end{eqnarray}}
\def\lb{\label}

\usepackage{graphicx}

\begin{document}

\begin{center}
\Large {\bf  Anomalies, entropy and boundaries}
\end{center}

\bigskip
\bigskip

\begin{center}
Dmitri~V.~ Fursaev$^1$,\ \ \  Sergey~ N.~ Solodukhin$^2$
\end{center}

\bigskip
\bigskip

\begin{center}
$^1$ {\it Dubna State University\\
     Universitetskaya str. 19
     141 980, Dubna, Moscow Region, Russia\\
      and  the Bogoliubov Laboratory of Theoretical Physics\\
  Joint Institute for Nuclear Research
  Dubna, Russia\\}
  
  \medskip

$^2$   {\it Laboratoire de Math\'ematiques et Physique Th\'eorique  CNRS-UMR
7350, \\
  F\'ed\'eration Denis Poisson, Universit\'e Fran\c cois-Rabelais Tours,  \\
  Parc de Grandmont, 37200 Tours, France}
 \medskip
\end{center}

\bigskip
\bigskip

\begin{abstract}
A relation between the conformal anomaly and the logarithmic term in the entanglement entropy is known to exist for CFT's in even dimensions. In odd dimensions the local anomaly  
and the logarithmic term in the entropy are absent.  
As was observed recently,  there exists
a non-trivial integrated anomaly if an odd-dimensional spacetime has boundaries.
We show that, similarly,  there  exists
a logarithmic term in the entanglement entropy when the entangling surface crosses the boundary of spacetime.
The relation of the entanglement entropy to the integrated conformal anomaly is 
elaborated for three-dimensional theories.
Distributional properties of intrinsic and extrinsic geometries of the boundary in the presence of conical singularities in the bulk are established. This allows one
to find contributions to the entropy that depend on the relative angle 
between the boundary and the entangling surface. 

\end{abstract}

\newpage

\section{Introduction}
\setcounter{equation}0

It is by now well known that for conformal field theories there is a relation between the conformal anomalies and the logarithmic term in the entanglement entropy
computed for a entangling surface $\Sigma$. This relation was first found in \cite{Fursaev:1994te} in the case when $\Sigma$ is a black hole horizon\footnote{The logarithmic term in $d=4$ entanglement entropy was discovered in \cite{Solodukhin:1994yz} for the Schwarzschild black hole.}.
The general structure of this relation,  see \cite{Ryu:2006ef},  can be obtained by applying the replica method in which the entropy is derived from the action by introducing a small
angle deficit at surface $\Sigma$ and then differentiating with respect to the deficit. The geometrical aspects of this technique were developed in \cite{Fursaev:1995ef}.
Black hole horizons, static or stationary, are characterized by vanishing extrinsic curvature. For a generic surface 
there appears a extra contribution, see  \cite{Solodukhin:2008dh}, to the logarithmic term in the entropy that is due to the extrinsic curvature of the surface.
Geometrically, these contributions originate from a squashed conical singularity as was shown in \cite{Fursaev:2013fta}.

In odd dimensions, the local conformal anomaly is absent since there is no invariant of
an odd dimension constructed from the Riemann curvature and its derivatives. 
Similarly, the logarithmic term in the entropy for any CFT in odd dimensions is absent. This is due to the fact that one cannot construct 
a geometric invariant of odd dimension from the bulk curvature and the extrinsic curvature of the entangling surface. Since the invariant in question should 
not depend on the choice of the vectors normal to the surface one always needs an even number of extrinsic curvatures to construct an invariant.

As became clear from the recent studies \cite{Fursaev:2015wpa}, \cite{Solodukhin:2015eca}, \cite{Herzog:2015ioa} the situation is different if the space-time in  question has boundaries.
First of all, the integrated conformal anomaly in even dimensions is modified by certain boundary terms as shows the explicit calculation \cite{Fursaev:2015wpa} in $d=4$ dimensions.
Then, quite surprisingly, the anomaly in odd dimensions occurs to be  non-trivial solely due to the boundary terms, as was suggested in \cite{Solodukhin:2015eca}.
The reason for this is simple: the boundary in this case is even-dimensional and the required conformal invariants of even dimension can be easily constructed 
 from the Riemann curvature and  the extrinsic curvature of the boundary.

In this paper we build on these findings and suggest that, similarly,  the logarithmic term in entanglement entropy is non-trivial  in odd dimensions provided the entangling surface
crosses the boundary of spacetime. The earlier study of the boundary entanglement is given in  \cite{Fursaev:2013mxa}.
For simplicity, we mainly focus on the $d=3$ case although the general argument is applicable to the entanglement entropy calculation in any odd-dimensional space-time
and we comment on the possible structure of the logarithmic term in higher dimensions.

Our results can be summarized as follows. The integrated anomaly in 
a conformal field theory in three dimensions
is solely determined by two  boundary conformal invariants $\chi_2$, the Euler number of 
the boundary, and $j$, a quadratic combination of the boundary extrinsic curvature, 
\be
\label{1.5}
{\cal A}=-a\chi_2+qj\, ,
\ee
and by two boundary charges $a$ and $q$ specific for the theory. The logarithmic term in the entanglement 
entropy which appears under
partition of the space-time by an entangling surface (curve) $\Sigma$ crossing the boundary
can be defined via the anomaly of the effective action by following \cite{Ryu:2006ef}.
We denote this term $s_{\mbox{\tiny{anom}}}$ and find
\be
s_{\mbox{\tiny{anom}}}=-a{\cal N}+q\sum_{k}f(\alpha_k)
\lb{ie-23}
\ee
where ${\cal N}$ is the number of points, where
$\Sigma$ crosses the boundary, and $\alpha_k$ are angles (at each of such points)
between a normal vector to the boundary and the tangent vector to $\Sigma$.

Derivation of (\ref{ie-23}) from (\ref{1.5}) requires one to consider 
(\ref{1.5}) on manifolds with conical singularities when the singular surface 
crosses the boundary. One of our new results 
is that quadratic combinations of the boundary extrinsic curvature tensor
behave as delta-functions in a way similar to components of the Riemann tensor \cite{Fursaev:1995ef}. The angle function $f(\alpha)$ can be calculated explicitly,
see (\ref{ie-f3}).

We also study the logarithmic term in the genuine entanglement entropy derived 
in a standard way from the divergent part of the effective action on a manifold with conical singularities. This term is denoted as $s$.  By analysing the case when $\Sigma$ is orthogonal to the boundary,
$\alpha_k=0$, we conclude that $s$ and $s_{\mbox{\tiny{anom}}}$ do not coincide, in general. 
The differences  appear due to the presence of the
non-minimal coupling.  This is a novel feature specific for 3-dimensional entropies.
This property does not appear in 4 dimensions \cite{Ryu:2006ef},\cite{Solodukhin:2008dh}.

The paper is organized as follows. In Sec. \ref{bc1} we compute the integrated conformal
anomaly in scalar and spinor CFT's in three dimensions. The Renyi entanglement entropy
in these theories for a 3D Euclidean space with two parallel boundary planes and $\Sigma$
orthogonal to the boundary planes is considered in Sec. \ref{bc2}. A relation between
this entropy and the integrated anomaly is studied in Sec. \ref{bc3}. 
Distributional properties of the boundary geometry due to conical singularities
when $\Sigma$ crosses the boundary plane under an arbitrary angle are analysed 
in Sections  \ref{bc4} and \ref{bc5}. In particular we present there new results on distributional properties of extrinsic curvatures.
Formula (\ref{ie-23}) and a relation of $s_{\mbox{\tiny{anom}}}$ to $s$ 
are considered in Sec. \ref{bc5a}.
A brief discussion of the logarithmic terms in the entropy in higher dimensions
and conclusions are in Sections \ref{bc6} and \ref{bc7}.

\section{Integrated conformal anomaly in $d=3$}\label{bc1}
\setcounter{equation}0

Boundary terms in conformal anomalies of effective actions of different conformal field theories have been studied recently in four dimensions \cite{Fursaev:2015wpa}, \cite{Solodukhin:2015eca}. The aim of these works was to establish the universal structure
of boundary terms and relation of boundary and bulk charges.

We consider an integral conformal anomaly of the effective 
action $W$ 
\begin{equation}\label{1.4a}
{\cal A}\equiv \partial_\sigma W[e^{2\sigma}g_{\mu\nu}]_{\sigma=0}=\int_{{\cal M}_d}
\sqrt{g}d^dx \langle T_\mu^\mu\rangle\, ,
\end{equation}
which is defined under scaling with a constant factor $\sigma$. The right hand side 
of (\ref{1.4a}) relates $\cal A$ to the trace anomaly and
holds on a closed manifold. In odd dimensions ($d=2k+1$) the local trace anomaly is 
zero. The integral anomaly, however, may be non-trivial in the case of boundaries.

We calculate the integral anomaly for a number of models for $d=3$ and comment on 
its generalizations to higher dimensions ($d=5$).  
For $d=3$  the anomaly is a pure boundary term (\ref{1.5}) composed of two conformal
invariants set on the boundary\footnote{See also \cite{Jensen:2015swa} for a related discussion in the presence of defects.}
\begin{equation}
\chi_2=\frac{1}{4\pi}\int_{\partial {\cal M}}\sqrt{H}d^2x~\hat{R}\, ,
\label{1.5-1}
\end{equation}
\begin{equation}\label{1.6}
j={1 \over 4\pi}\int_{\partial {\cal M}}\sqrt{H}d^2x~
\mbox{Tr}~\hat{K}^2\, .
\end{equation}
Here $\chi_2$ is the Euler number of $\partial {\cal M}$, $\hat{R}$ is the curvature on the boundary,  $a$ and $q$ are boundary charges which depend on the model under consideration. Conformal invariance of $j$ in (\ref{1.6}) follows from the fact
that $\hat{K}_{\mu\nu}$ transforms homogeneously under conformal transformations.

We use the following notations: $g_{\mu\nu}$ is the metric of the background manifold 
$\cal M$,  the metric induced on the boundary $\partial {\cal M}$ of $\cal M$ is
$H_{\mu\nu}=g_{\mu\nu}-N_{\mu}N_\nu$, 
$N^\nu$ is a unit outward pointing normal vector to $\partial {\cal M}$, $K_{\mu\nu}=H_{\mu}^\lambda H_{\nu}^\rho N_{(\lambda;\rho)}$ is the extrinsic curvature tensor
of $\partial {\cal M}$, 
$\hat{K}_{\mu\nu}=K_{\mu\nu}-H_{\mu\nu}K/2$ is a traceless part of $K_{\mu\nu}$.
Also we put 
$\mbox{Tr}K^m=K^{\mu_1}_{\mu_2}...K^{\mu_p}_{\mu_1}$, $K=\mbox{Tr}K$.

In three dimensions we have 3 non-interacting conformal field theories: massless scalar fields with a conformal coupling (either with the Dirichlet or with the conformal Neumann condition) and a massless
spin 1/2 field with mixed boundary conditions. The boundary charges for these models are listed in Table \ref{t1}.  

\begin{table}
\renewcommand{\baselinestretch}{2}
\medskip
\caption{Charges in the anomaly of the effective action}
\bigskip
\begin{centerline}
{\small
\begin{tabular}{|c|c|c|c|c|c|}
\hline
$\mbox{Theory}$  & $a$ & $q$  & $\mbox{boundary
condition}$   \\
\hline
$\mbox{real scalar}$  & ${1\over 96}$ 
 &  ${1 \over 64}$   & $\mbox{Dirichlet}$ \\
\hline
$\mbox{real scalar}$  & $-{1\over 96}$ 
 &  ${1 \over 64}$   & $\mbox{Robin}$ \\
\hline
$\mbox{Dirac spinor}$ 
 &  $0$ &
 ${1 \over 32}$ & $\mbox{mixed}$  \\
\hline
\end{tabular}}
\bigskip
\renewcommand{\baselinestretch}{1}
\end{centerline}
\label{t1}
\end{table}

Boundary charge $a$ is not positive definite: it changes the sign depending on the type of the boundary condition.
On the other hand, the boundary charge $q$ appears to be positive and independent of the boundary conditions. 
In a renormalization in which for a scalar field the value of this charge is $q=1$ 
it takes value $q=2$ for a Dirac fermion. This behavior suggests that $q$ may
represent the number of degrees of freedom in the quantum field theory.
The charges for the spinor theory are just a sum of the charges of two scalar CFT's
with the Dirichlet and Newmann boundary conditions.

In the scalar case $a$ and $q$ have been computed in \cite{Solodukhin:2015eca}.
For completeness we comment on computations of charges in Table \ref{t1}.

We use the relation between the anomaly and the heat coefficient 
of a Laplace operator $\Delta=-\nabla^2+X$ for the corresponding $d$-dimensional conformal theory,
\begin{equation}\label{1.4b}
{\cal A}= \eta A_d~~,
\end{equation}
where $\eta=+1$, for Bosons, and $\eta=-1$, for Fermions. The heat coefficients for the asymptotic expansion of the heat kernel of $\Delta$ are defined as
\begin{equation}\label{1.1}
K(\Delta;t)=\mbox{Tr}~e^{-t\Delta}\simeq \sum_{p=0} A_p(\Delta)~t^{(p-d)/2}~~,~~t\to 0~~.
\end{equation}

If the classical theory is scale invariant the heat coefficient $A_d$ is 
a conformal invariant, see e.g. \cite{Fursaev:2011zz}. Therefore $A_d$ can be represented
as a linear combination of conformal invariants constructed of the geometrical characteristics of $\cal M$, $\partial {\cal M}$ and embedding of $\partial {\cal M}$ in
$\cal M$. 

We are interested in   $d=3$ case.  The heat coefficient in this case is a pure boundary term. On dimensional grounds $A_3$ can be written
in terms of 2 invariants $\chi_2$ $j$. Thus, one comes to (\ref{1.5}). 
Explicit values of the charges can be found by using results for the heat kernel coefficients
collected in \cite{Vassilevich:2003xt}. The boundary conditions are
\begin{equation}\label{1.20}
(\nabla_N-S)\Pi_+\phi=0~~,~~\Pi_-\phi=0~~~,
\end{equation}
where $\nabla_N=N^\mu\nabla_\mu$, $\Pi_\pm$ are corresponding projectors,
$\Pi_++\Pi_-=1$, definition of $S$ coincides with \cite{Vassilevich:2003xt}.
Some general formulas for the heat kernel coefficients (see \cite{Vassilevich:2003xt}) in dimension $d$ are
\be
A_1={1 \over 4 (4\pi)^{\frac{d-1}{2}}}\int_{\partial {\cal M}}\sqrt{H}d^{d-1}x\tr~\chi\, ,
\lb{AA1}
\ee
\be
A_3={1 \over 384 (4\pi)^{\frac{d-1}{2}}}\int_{\partial {\cal M}}\sqrt{H}d^{d-1}x~
\mbox{Tr}~\left[-96 \chi X+16 \chi R -8\chi R_{\mu\nu}N^\mu N^\nu\right.\nonumber \\
\left.+(13\Pi_+-7\Pi_-)K^2+(2\Pi_+ +10\Pi_-)\mbox{Tr}K^2 +
96SK +192 S^2-12\chi_{:a}\chi_{:a}\right]~~.
\lb{1.11}
\ee
Here we use 'flat' indices $a,b$ in the tangent space 
to the boundary, $\chi=\Pi_+ -\Pi_-$.

For $d=3$ in the case of a conformal scalar field  $X=R/8$. For the Dirichlet condition $\Pi_+=0$, $\chi=-1$. Conformally invariant scalar Robin condition requires
$S=-K/4$, $\Pi_+=1$, $\chi=1$. 

In case of a massless Dirac field $\psi$ the operator is
$\Delta^{(1/2)}=(i\gamma^\mu\nabla_\mu)^2$. The boundary conditions
are mixed ones,
\begin{equation}\label{2.10-dd}
\Pi_- \psi\mid_{\partial {\cal M}}=0~~,~~
(\nabla_N+K/2)\Pi_+ \psi\mid_{\partial {\cal M}}=0~~,
\end{equation}
where $\Pi_\pm=\frac 12 (1\pm i\gamma_\ast N^\mu \gamma_\mu)$, and 
$\gamma_\ast$ is a chirality gamma matrix. Therefore, $X=R/4$, $S=-\Pi_+K/2$. 

Results of Table \ref{t1} easily follow from (\ref{1.11})
if we use the Gauss-Codazzi identities and relations, $\mbox{Tr}\Pi_\pm=1$, 
$\chi_{:a}=2\Pi_{+:a}=i\gamma^b\gamma_\ast K_{ba}$, 
$\mbox{Tr}~\chi_{:a}\chi_{:a}=2\tr K^2$.

\section{Entanglement Renyi entropy in $d=3$}\label{bc2}
\setcounter{equation}0

We consider the entropy of entanglement 
and, more generally, entanglement Renyi  entropy $S^{(n)}$. 
The order of the entropy $n$
is a natural number, entanglement entropy corresponds to value $n=1$. The entropy is assumed to appear 
under partition of the system into different parts by an entangling surface 
$\Sigma$. In $d=3$ $\Sigma$ is a curve. 
The Renyi entropy in three dimensions is expected to depend on the UV cutoff 
parameter as follows:
\begin{equation}\label{1.5e}
S^{(n)}\simeq c(n)L \Lambda + \ln (\mu\Lambda) \sum_P s(n,\alpha_k) \, ,
\end{equation}
where $L$ is the length of $\Sigma$, $\Lambda$ is the UV cutoff and $\mu$ is a typical scale of the system, $c(n)$ is some polynomial of $n$. The aim of our analysis is the logarithmic term of the entropy in (\ref{1.5e}). This term, as we shall see, comes  from the points where the entangling surface $\Sigma$ crosses the boundary. We denote this set of points as $P=\Sigma   \cap {\partial{\cal M}}=\cup_k p_k$. The sum in (\ref{1.5e}) goes over all such points $p_k$.
Coefficient functions $s(n,\alpha_k)$ are polynomials of $n$. Their conformal invariance
in a CFT allows dependence on the angles $\alpha_k$ between $N$ and tangent vectors to $\Sigma$ at those boundary points.

The question which we address in this paper is the relation between the integrated conformal 
anomaly of the effective action and
the logarithmic term
in (\ref{1.5e}). We study the models introduced in Sec. \ref{bc1}.

The logarithmic part of the entropy in (\ref{1.5e}) can be derived 
from the ultraviolet part of the effective action by setting
the action on manifolds with conical singularities. The method 
is described, e.g. in \cite{Fursaev:2012mp},
and yields the formula:
\begin{equation}\label{3.2}
s(n)=\eta{nA_3(1)-A_3(n) \over n-1}~
\end{equation}
(the zero modes in (\ref{3.2}) are ignored). Here $A_3(n)$ is the heat coefficient
of the corresponding Laplacian on a manifold ${\cal M}_n$ which appears
in the replica method of the entropy computations. ${\cal M}_n$ is constructed from $n$ copies of the spacetime manifold $\cal M$ by a 'cut and glue' procedure. The details of this
procedure are not relevant for our purposes.

The heat coefficients $A_3(n)$ when $\Sigma$ meets the boundary under an arbitrary 
angle are unknown. So we derive them for configurations when $\Sigma$ and the boundary
are orthogonal. We suppose that the spacetime ${\cal M}$ has 
the structure $R^2\times L$, where $L$ is an interval with 2 end points.
The space $\cal M$ is flat and the operators reduce to simple Laplacians.

The boundary $\partial {\cal M}$ consists of two planes (located at end points 
of $L$). The entangling surface $\Sigma$ is a interval 
(identical to $L$).  $\Sigma$ touches $\partial {\cal M}$  at its two end 
points, $p_1$ and $p_2$. The replica method introduces manifolds ${\cal M}_n={\cal C}_n\times L$,
where ${\cal C}_n$ is a two-dimensional cone with an opening angle $2\pi n$.
The heat kernel coefficient $A_3(n)$ on ${\cal M}_n$ is a product 
\be
A_3(n)=A_2({\cal C}_n)\times A_1(L)\, ,
\lb{A3}
\ee
where 
\be
A_2({\cal C}_n)=\eta \frac{1}{12n}(1-n^2)
\lb{A2}
\ee
is the heat kernel coefficient for a Laplace operator for the considered models on a two-dimensional cone, and 
\be
A_1(L)=\frac{1}{4}\sum_{P_k}\tr ~\chi
\lb{A1}
\ee
is the heat coefficient on one-dimensional interval $L$, with two ends $p_1$ and $p_2$, see Eq.(\ref{AA1}).

The result for the entropy is :
\begin{equation}\label{3.3}
s(n)=-4a{n+1 \over n}~,\ \  s=-8a~~,
\end{equation}
where $a$ are defined in Table \ref{t1}.
Here $s=s(1)$ is the logarithmic term in the entanglement entropy. 

No logarithmic terms in the entropy appear for spin 1/2 fields, since $\tr \, \chi=0$.
For scalar fields each point of the boundary, where it meets the entangling surface,
yields a contribution to the logarithmic term (\ref{3.3}) 
equal $-4a$, $a=\pm 1/96$. The extra factor 2 in (\ref{3.3}) follows
from the fact that there are 2 such points. In general, if there are $\cal N$
such points, the logarithmic term is $s=-4{\cal N} a$.

\section{Entropy from the integrated conformal anomaly, and non-minimal coupling}\label{bc3}
\setcounter{equation}0

The logarithmic terms in entropy of entanglement can be alternatively  derived from
the anomaly of the effective action, as has been shown in \cite{Ryu:2006ef}. 
This definition for the integrated anomaly looks as follows:
\begin{equation}\label{4.2.0}
s_{anom}=\lim_{n\to 1}{n{\cal A}-{\cal A}(n) \over n-1}~~~.
\end{equation}
Here ${\cal A}(n)$ is the integrated anomaly (\ref{1.5}) taken on the corresponding replicated manifolds ${\cal M}_n$ glued from $n$ copies of $\cal M$. Formula (\ref{4.2.0})
requires definitions of invariants $\chi$ and $j$ on the boundary
$\partial {\cal M}_n$ in case when $\Sigma$ crosses $\partial {\cal M}_n$.

We show that (\ref{4.2.0}) results in (\ref{ie-23}). To simplify the analysis in this Section we again consider the case of the entangling surface orthogonal to the boundary. Conical singularities result in delta-function contributions
in the scalar curvature of $\partial {\cal M}_n$. Thus,  we expect that $\chi$ should produce 
a non-trivial contribution in (\ref{4.2.0}), since the Euler characteristics
in two dimension  is determined by the curvature scalar. As for another invariant $j$, Eq. (\ref{1.6}),  we demonstrate in next Section that $\tr \hat{K}^2$ can be non-trivial only when $\Sigma$ is not orthogonal to the boundary.

Let us see what happens with the integrated anomaly (\ref{1.5}) by using
example of the previous Section,
${\cal M}_n={\cal C}_n\times L$. 
Since invariant $q$ does not contribute at the conical singularity then 
\begin{equation}\label{4.1}
{\cal A}(n)=-a\chi_2[\partial {\cal M}_n]=-2a(1-n)\, .
\end{equation}
Coefficient 2 in the r.h.s. of (\ref{4.1}) corresponds to two components of the boundary.
For the configuration in question we put $\chi_2[\partial {\cal M}_n]=2(1-n)$, where
we took into account that each conical singularity adds an extra $4\pi(1-n)$ 
to the integral of the curvature. The logarithmic term in the entropy obtained 
via (\ref{4.2.0}) is
\begin{equation}\label{4.2.0a}
s_{anom}=-2a\, .
\end{equation}
Formula (\ref{4.2.0a}) is a particular case of (\ref{ie-23}) for ${\cal N}=2$ and
$\alpha_k=0$.

In case of scalar theory (\ref{4.2.0a})  contradicts the straightforward  derivation 
of the logarithmic term, $s=-8a$, see (\ref{3.3}).
The explanation of this contradiction is as follows. In a scalar CFT the 
corresponding Laplace operator $\Delta=-\nabla^2+X$ includes the non-minimal coupling
with the curvature, since the conformal invariance requires that $X=R/8$.
Computation of $s_{anom}$ implies that all curvatures including 
those in the non-minimal couplings  should acquire delta-function
terms. Opposite to that, calculation of $s$ treats $X$ as a non-singular term in the Laplacian.

Contribution of the non-minimal coupling
to the heat coefficient is, see (\ref{1.11}),
\begin{equation}\label{4.2}
A_3^{nc}={1 \over 384 (4\pi)}\int_{\partial {\cal M}}\sqrt{H}d^2x~
\mbox{Tr}~\left[-12 \chi R\right]\, .
\end{equation}
If conical singularities are allowed to contribute in (\ref{4.2}) one finds
\begin{equation}\label{4.2cs}
A_3^{nc}(n)=nA_3^{nc}(1)-{1-n \over 32 }
\sum_k\mbox{Tr}~\chi_k\, .
\end{equation}
Notice that in (\ref{4.2}) the bulk curvature $R$ appears in the integral over 
the boundary.  For the orthogonal configuration this makes no problem since the delta-function in $R$
is in fact the delta-function normalized to unity with respect to the integration over 
the boundary. Formula (\ref{4.2cs}) results in the following addition to the entropy:
\begin{equation}\label{4.3}
s_{nc}=\lim_{n\to 1}{nA_3^{nc}(1)-A_3^{nc}(n) \over n-1}
=-{1 \over 32 }\sum_k\mbox{Tr}~\chi_k=6a\, .
\end{equation}
Now one can check that
\begin{equation}\label{4.4}
s_{anom}=s+s_{nc}\, .
\end{equation}

The effect of the non-minimal couplings is a new feature in three dimensional 
theories. This feature is not known in four dimensions. There, despite the presence of the non-minimal coupling in the scalar conformal operator, 
there is a complete agreement between the direct entropy calculation and the calculation via the conformal anomaly.

\section{Conical singularity and induced geometry on the boundary}\label{bc4}
\setcounter{equation}0

So far our analysis has been restricted to cases where the boundary and the entangling surface
are orthogonal.
In this section 
we study distributional properties of the boundary geometry provided the singular surface
crosses the boundary at an arbitrary angle. To this aim it is 
useful to consider a simplest three-dimensional set up in which the 3-dimensional space-time $\cal M$, its boundary and the entangling surface are flat. We choose Cartesian coordinates  
$(x^0,x^1,x^2)$ on $\cal M$ where the metric is
\be
ds^2=dx_0^2+dx_1^2+dx_2^2\, .
\lb{c1}
\ee
The entangling surface $\Sigma$ is supposed to be a plane $x_1=0$, while the boundary $\partial {\cal M}$ is an arbitrary plane parallel to $x^0$
\be
x_2=a x_1\, , \ \  a=\tan\alpha\,  , \ \ \cos\alpha=N^\mu l_\mu\, ,
\lb{c2}
\ee
where $N^\mu$ is vector normal to the boundary $\partial {\cal M}$ and $l^\mu$ is vector tangent to the entangling surface $\Sigma$. 
The boundary $\partial{\cal M}$ and the entangling surface intersect at point $P$.
One sees that  $(\pi/2-\alpha)$ is the angle with which $\Sigma$ intersects the boundary $\partial {\cal M}$ on the hypersurface of constant time $x_0$.

To introduce ${\cal M}_n$ we switch to polar coordinates in 
the metric of $\cal M$:
\be
x_0=\rho\sin n\phi \, \ \ x_1=\rho\cos n\phi
\lb{c3}
\ee
and assume that $\phi$ has period $2\pi$. ${\cal M}_n$ has a conical singularity 
at $\rho=0$ with the angle surplus $2\pi(1-n)$.  

Since the singular surface crosses the boundary we expect some sort of conical
singularities on the boundary. We first study intrinsic geometry on the boundary 
by using two independent methods in the limit $n\to 1$ (by assuming, as usual,
that such a limit can be done in final integral expressions). 

It is instructive to start with the analysis of the distributional properties of the scalar
curvature of the boundary by using a regularization procedure \cite{Fursaev:1995ef}. The subtle point is that we want to regularize 
the boundary geometry by regularizing the bulk geometry.
The regularized bulk metric takes the form
\be
ds^2=f_n(\rho)d\rho^2+n^2\rho^2d\phi^2+dx_2^2\, ,  \ \  f_n(\rho)=\frac{\rho^2+n^2b^2}{\rho^2+b^2}\, ,
\lb{c4}
\ee
where $b$ is a regularization parameter to be taken to zero at the end.
The regularity of (\ref{c4}) at $\rho=0$ can be seen in coordinates
$y_1=\rho \cos \phi$, $y_2=\rho \sin \phi$ which near $\rho=0$ behave as Cartesian
coordinates.

To find regularized geometry induced on the boundary $\partial {\cal M}$ 
one  should take (\ref{c4}) on some embedding equation which does not spoil
analyticity at $\rho=0$. Equation (\ref{c2}) in the polar coordinates takes the form:
\be
x_2=a\rho\cos n\phi\, , \ \ a=\tan\alpha\, 
\lb{c41}
\ee
and cannot be used since it does not have the analytic behaviour 
due to the presence of $\cos n\phi$. The problem is similar to manifolds with squashed conical singularities considered in \cite{Fursaev:2013fta} and 
it is cured by changing embedding equation (\ref{c41})
to
\be
x_2=a~c^{1-n}\rho^n\cos n\phi\, , \ \  n>1~~,
\lb{ie-18}
\ee 
where $c$ is an arbitrary parameter which has the dimension of a length. Coordinate
$x_2$ in (\ref{ie-18}) is an analytic function with respect to Cartesian-like coordinates $y_1,y_2$.

One can now compute the integral of the scalar curvature $\hat{R}$ of the 
regularized boundary geometry over some domain around $\rho=0$,
say a disk of a finite radius $\rho_0$ with the center at $\rho=0$. 
Calculations can be done in two steps. First, one computes
the integral of $\hat{R}$ along $\rho$ with the condition that $f(\rho=0)=n^2$ and 
$f(\rho_0)\rightarrow 1$, when $b\rightarrow 0$.
With these conditions the integral does not depend on the concrete form of $f(\rho)$. 

The integration over angle $\phi$ is complicated. To perform it  
we relax the condition that $n$ is a natural number and consider the limit $n\to 1$, which 
is actually what we are interested in.  By decomposing the integral
in powers of $(n-1)$ we obtain
\be
\int_0^{2\pi}d\phi \int_0^{\rho_0}d\rho \sqrt{H} \hat{R}=\frac{4\pi}{\sqrt{1+a^2}}(1-n)+O(1-n)^2\, ,
\lb{c8}
\ee
where the higher order terms are functions of $a$ (vanishing identically if $a=1$).
This means that $\hat{R}$ has a delta-function behaviour at point $\rho=0$  
\be
\hat{R}\simeq 4\pi\cos\alpha ~(1-n)\hat{\delta}_P\, ,~~n\to 1~~,
\lb{c9}
\ee
where $\hat{\delta}_P$ is normalized with respect to the integration measure on 
$\partial {\cal M}$.

If the entangling surface and the boundary are orthogonal, 
$\alpha=0$, the boundary conical singularity at $P$ results in the usual
addition $4\pi(1-n)$ to the curvature integral.
For arbitrary $\alpha$ we find an extra factor $\cos(\alpha)$.
If the entangling surface touches the boundary at zero angle $(\alpha=\pi/2)$ the conical singularity on the boundary effectively disappears.

We now obtain (\ref{c9}) by an alternative method based on some simple geometrical
arguments. Note that in the considered geometrical setup (Eqs. (\ref{c1}), (\ref{c2})) 
there is a rotational symmetry around the conical singularity. Instead of studying $n$ replicas of $\cal M$ with conical singularities 
on the entangling line $x_0=x_1=0$ (and the surplus of the conical angle) we can consider 
a manifold ${\cal M}_\beta$ with conical singularities on $x_0=x_1=0$ and the length of a unit radius circle 
around the singularity equal $2\pi \beta$. As a result of the rotational symmetry, 
$\beta$ can be an arbitrary positive parameter. We choose $0<\beta<1$ which corresponds to
the conical angle deficit. 

In this case  the orbifold ${\cal M}_\beta$ can be obtained by cutting a part of the  
3D Euclidean space between two planes which intersect at the line $x_0=x_1=0$ 
under angle $\delta=2\pi(1-\beta)$, $\delta$ being the angle deficit of the cone. The boundary $\partial {\cal M}_\beta$ of ${\cal M}_\beta$
is a 2-plane with a conical singularity. When the two planes intersect the boundary 
they produce two lines under the angle $\delta_b=2\pi(1-\beta_b)$, which can be determined from the
relation
\be
\tan (\delta_b)=\cos\alpha ~\tan (\delta) \, .
\lb{ie-01}
\ee
We see that the angle deficit on the boundary is different from the one in the bulk.
At small angle deficits ($\beta \to 1$)  the relation (\ref{ie-01}) takes a simple form
\be
2\pi (1-\beta_b)\simeq  2\pi(1-\beta)\cos\alpha\, .
\lb{ie-02}
\ee

The boundary metric on $\partial {\cal M}$ can be written as
\be
dl^2=dx_0^2+(1+a^2)dx_1^2=dx_0^2+{dx_1^2 \over \cos^2\alpha}=dx_0^2+d{\bar x}_1^2 \, .
\lb{ie-1}
\ee
The same metric can be used on  $\partial {\cal M}_\beta$ if we introduce a set of polar coordinates
\be
x_0=\bar{\rho}\sin(\bar{\phi}\beta_b)\, \ \ {\bar x}_1=\bar{\rho}\cos(\bar{\phi} \beta_b)
\lb{ie-2}
\ee
where $\bar{\phi}$ ranges from 0 to $2\pi$.
The scalar curvature for the metric induced on the boundary, therefore, acquires
the following contribution at small angle deficits: 
\be
\hat{R}=4\pi(1-\beta_b)\hat{\delta}_P \simeq 4\pi \cos\alpha (1-\beta)\hat{\delta}_P\, .
\lb{ie-3b}
\ee
This results agrees with Eq. (\ref{c8}) obtained by using the regularization procedure.
We use these results in next Sections.

\section{Extrinsic geometry of conical singularity on the boundary}\label{bc5}
\setcounter{equation}0

One of the results of this paper is that not only intrinsic geometry but also
geometrical characteristics related to its embedding in a higher dimensional space may have distributional properties. Let us demonstrate this for invariants quadratic in extrinsic curvatures, $K^2$, $\tr K^2$, by using the model introduced in the previous section.

Regularization (\ref{c4}), (\ref{ie-18}) makes the extrinsic curvature tensor regular at $\rho=0$ and integrable provided $n>1$. If instead of regularized 
embedding equation (\ref{ie-18}) one used equation (\ref{c41}) the components of the extrinsic curvature tensor had a power law singularity in $\rho$ at $\rho=0$.
 
Since for $n=1$ the extrinsic curvature vanishes one would expect 
that integrals of $K^2$, $\tr K^2$ yield results which are at least quadratic in $(n-1)$. 
A careful analysis shows that it is not so. The integration over a disk of radius $\rho_0$  produces 
terms $\rho_0^{2n-2}/2(n-1)= 1/2(n-1)+O(1)$. So that in the limit $n\rightarrow 1$ 
this pole cancels one power of $(n-1)$ and transforms $(n-1)^2$ term into $(n-1)$ term. This is exactly the same mechanism which happens in the case
of the squashed cones considered in \cite{Fursaev:2013fta}.
With these remarks in the limit of small angle deficit ($n\to 1$) 
one obtains the following 
expressions:
\be
\int_{\partial {\cal M}_n} K^2\simeq \int_{\partial {\cal M}_n} \mbox{Tr}~ K^2\simeq 
8\pi (1-n) f(\alpha)~~,
\lb{ie-f1}
\ee
\be
j\simeq 
(1-n) f(\alpha)~~,
\lb{ie-f2}
\ee
\be
f(\alpha)=-\frac{1}{32} \frac{\sin^2\alpha}{\cos\alpha}(1+2\cos^2\alpha+5\cos^4\alpha)
\lb{ie-f3}~~.
\ee
Invariant $j$ is defined in (\ref{1.6}).

Our next step is to study the effect of conical singularities in quadratic combinations
of $K_{\mu\nu}$ of a general form and check the consistence of expressions (\ref{ie-f1}).
We continue with a 3D manifold ${\cal M}_\beta$ with a rotational symmetry 
around conical singularities and return to an arbitrary angle $\beta$ 
around the singular points. Components of the Riemann tensors on ${\cal M}_\beta$ and 
its boundary $\partial {\cal M}_\beta$, respectively, are
\be
R_{\mu\nu\lambda\rho}=2\pi(1-\beta)\left(G_{\mu\lambda}G_{\nu \rho}-
G_{\mu\rho}G_{\nu \lambda}\right)\delta_P\, ,
\lb{ie-5}
\ee
\be
\hat{R}_{\mu\nu\lambda\rho}=2\pi(1-\beta_b)\left(H_{\mu\lambda}H_{\nu \rho}-
H_{\mu\rho}H_{\nu \lambda}\right)
\hat{\delta}_P\, .
\lb{ie-6}
\ee
Here $G_{\mu\nu}=(p_i)_\mu(p_i)_\nu$ and $p_i$ is a pair of orthonormalized vectors
orthogonal to the entangling line in $\cal M$, while $H_{\mu\nu}=(\hat{p}_i)_\mu(\hat{p}_i)_\nu$, where $\hat{p}_i$ is a pair 
of orthonormalized vectors at the conical singularity 
in the tangent space of $\partial {\cal M}$. In fact, $H_{\mu\nu}$ is an induced metric 
on $\partial {\cal M}$.
In (\ref{ie-5}), (\ref{ie-6}) we also denote $\delta_P$ and 
$\hat{\delta}_P$ normalized delta-functions on ${\cal M}_\beta$ and 
$\partial {\cal M}_\beta$, respectively. The delta-functions in the limit $\beta\to 1$
are related as

\be
\hat{\delta}_P=\delta(x_0)\delta({\bar x}_1)=\cos\alpha~ 
\delta(x_0)\delta(x_1)=\cos\alpha \delta_P, ,
\lb{ie-3}
\ee
where we used the fact that ${\bar x}_1=x_1/\cos\alpha$.  

One can suggest the following general structure:
\be
K_{\mu\nu}K_{\lambda\rho}=2\pi(1-\beta_b)\left(a_1 H_{\mu\nu}H_{\lambda\rho}+
a_2(H_{\mu\lambda}H_{\nu \rho}+H_{\mu\rho}H_{\nu\lambda})\right)
\hat{\delta}_P\, ,
\lb{ie-6b}
\ee
where $a_1$ and $a_2$ are unknown functions of $\alpha$. Eq. (\ref{ie-6b}) reflects
required symmetries under permutations of indices.
Another property of Eq. (\ref{ie-6b})  is that it takes into account $O(2)$ symmetry
associated with the rotation of the pair of vectors $\hat{p}_i$. This property can be seen
if one rewrites (\ref{ie-6b})  in the basis $\hat{p}_i$
\be
K_{ij}K_{mn}=2\pi(1-\beta_b)\left(a_1 \delta_{ij}\delta_{mn}+
a_2(\delta_{im}\delta_{jn}+\delta_{in}\delta_{jm})\right)
\hat{\delta}_P\, ,
\lb{ie-6c}
\ee
where $K_{ij}=\hat{p}^\mu_i \hat{p}^\nu_j K_{\mu\nu}$.

Consider restrictions imposed by the Gauss-Codazzi
equations on Eq. (\ref{ie-6b}). 
For smooth geometries one has
\be
K_{\mu\rho}K_{\nu\lambda}-K_{\mu\lambda}K_{\nu\rho}=R^{\Vert}_{\mu\nu\lambda\rho}-\hat{R}_{\mu\nu\lambda\rho}\, ,
\lb{ie-4}
\ee
where $R^{\Vert}_{\mu\nu\lambda\rho}$ are the components of the Riemann
tensor in the bulk projected on the tangent space to  $\partial {\cal M}$.
Let us study the right hand side of (\ref{ie-4}) by using expressions 
(\ref{ie-5}), (\ref{ie-6}) for singular parts. In the limit of a small angle deficit
$$
R^{\Vert}_{\mu\nu\lambda\rho}-\hat{R}_{\mu\nu\lambda\rho}\simeq
$$
\be
2\pi{(1-\beta) \over \cos\alpha}\left(G^{\Vert}_{\mu\lambda}G^{\Vert}_{\nu \rho}-
G^{\Vert}_{\mu\rho}G^{\Vert}_{\nu \lambda}-\cos^2\alpha(H_{\mu\lambda}H_{\nu \rho}-
H_{\mu\rho}H_{\nu \lambda})
\right)\hat{\delta}_P\, 
\lb{ie-7}
\ee
where we took into account (\ref{ie-3}).
We need to calculate $G^{\Vert}_{\mu\nu}=H_\mu^\lambda H_\nu^\rho G_{\lambda\rho}$. Let $l$ be 
a unit vector tangent to the entangling line. One has
\be
G_{\mu\nu}=\delta_{\mu\nu}-l_\mu l_\nu\, .
\lb{ie-8}
\ee
Therefore,
\be
G^{\Vert}_{\mu\nu}=H_{\mu\nu}-l^{\Vert}_\mu l^{\Vert}_\nu\, ,
\lb{ie-9}
\ee
where $l^{\Vert}_\mu=H_\mu^\lambda l_\lambda$.

Let us choose the basis $\hat{p}_i$ such that $\hat{p}_2$ is directed along the $x^0$ 
axis, while $\hat{p}_1$ is directed along the $x^1$. Then vector $l^{\Vert}$ is directed 
along $\hat{p}_1$, and one can easily see that  $l^{\Vert}=\sin\alpha~ \hat{p}_1$.
With these definitions,
$$
G^{\Vert}_{\mu\lambda}G^{\Vert}_{\nu \rho}-
G^{\Vert}_{\mu\rho}G^{\Vert}_{\nu \lambda}-\cos^2\alpha(H_{\mu\lambda}H_{\nu \rho}-
H_{\mu\rho}H_{\nu \lambda})=\sin^2\alpha\left(H_{\mu\lambda}H_{\nu \rho}-
H_{\mu\rho}H_{\nu \lambda}+ \right.
$$
\be
\left.H_{\mu\rho}(\hat{p}_1)_\nu(\hat{p}_1)_\lambda +
H_{\nu\lambda}(\hat{p}_1)_\mu(\hat{p}_1)_\rho-
H_{\mu\lambda}(\hat{p}_1)_\nu(\hat{p}_1)_\rho-
H_{\nu\rho}(\hat{p}_1)_\mu(\hat{p}_1)_\lambda\right)\equiv 0\, ,
\lb{ie-10}
\ee
Therefore, in the limit of a small angle deficit we have the following condition
for the singular parts:
\be
K_{\mu\rho}K_{\nu\lambda}-K_{\mu\lambda}K_{\nu\rho}=0\, ,
\lb{ie-11}
\ee
which should hold up to terms quadratic in $\beta-1$. It follows immediately 
that expressions (\ref{ie-f1}) obtained via the prescribed regularization procedure
are consistent with (\ref{ie-11}). Another consequence is that $a_1=a_2$ in  (\ref{ie-6b}), and in the limit $\beta\to 1$ one can write
\be
K_{\mu\nu}K_{\lambda\rho}\simeq \pi(1-\beta)f\left(H_{\mu\nu}H_{\lambda\rho}+
H_{\mu\lambda}H_{\nu \rho}+H_{\mu\rho}H_{\nu\lambda})\right)
\hat{\delta}_P~~~,
\lb{ie-12}
\ee
where $f=f(\alpha)$ is defined in (\ref{ie-f3}).

\section{A general expression for the entropy}\label{bc5a}
\setcounter{equation}0

We can now provide arguments in support of formula (\ref{ie-23}) for a general
geometric configuration. We use definition (\ref{4.2.0}) for entropy $s_{anom}$
derived from the integrated conformal anomaly (\ref{1.5}) and definitions of 
invariants  (\ref{1.5-1}), (\ref{1.6}). 

The expression at $q$-charge in (\ref{ie-23}) comes out immediately from (\ref{ie-f2}). The term associated to $a$-charge requires one to
know  topological characteristic $\chi$ in (\ref{4.2.0}). 
Despite the fact that we find a dependence on angle $\alpha$ in the formula for the boundary curvature (\ref{c9})
this dependence disappears in the Euler characteristics of the boundary. This is clearly the case for the Euler number of a disk 
containing a conical singularity. Indeed, the regularization as in Section 5 does not change the Euler number. Therefore, in the limiting
procedure the number for a disk remains to be $1$  and thus does not depend neither on the number of replicas $n$ nor on the angle $\alpha$.
To see what happens for compact boundaries
one can use again a simple set up. Suppose that ${\cal M}$ is a flat Euclidean
three dimensional space. Choose $\partial {\cal M}$  as a cylinder  
with the axis
along the entangling surface $\Sigma$, which is, say, $x^2$ axis, as in the examples considered
before. We suppose that the cylinder has the length $L$ and some radius. The 
size of cylinder
is restricted by two parallel planes going along $x^0$ and having an angle $\alpha$
with the axis $x^2$.  The topology of the complete boundary is that of a sphere $S^2$. Hence
$\chi[\partial {\cal M}]=2$. Since $\Sigma$ is along $x^2$ axis, ${\cal M}_n$ are
obtained from $n$ copies of $\cal M$ when cuts are made by a half plane which end on 
$x^2$. By this construction $\partial {\cal M}_n$ has the same topology as 
$\partial {\cal M}$ regardless
the value of $\alpha$. Therefore,
\begin{equation}\label{chi1}
{n\chi[\partial {\cal M}]-\chi[\partial {\cal M}_n]\over n-1}=2~~~,
\end{equation}
and one comes to (\ref{ie-23}) with ${\cal N}=2$. One can notice that 
each boundary end point of $\Sigma$ in this example
yields addition $-a$ to the $a$-term in (\ref{ie-23}). This can be extended to
an arbitrary number $\cal N$ of boundary points.

We now turn to the entropy $s$ which is defined by (\ref{3.2}) in the limit $n\to 1$.
Formula (\ref{3.2}) can be used if heat coefficients $A_3(n)$ are known for an arbitrary
angle between the singular surface and the boundary. Since straightforward derivations
of $A_3(n)$ is this case are absent one can apply results of Sections \ref{bc4} and \ref{bc5}
to formula (\ref{1.11}) by assuming this can be done at small angle deficits.
For spin 1/2 field it yields $s=s_{anom}$. For the scalar theory $s\neq s_{anom}$
since there are non-minimal couplings. To calculate $s$ we can proceed as in Sec. \ref{bc3}
and write, by taking into account  Eqs. (\ref{4.3}),(\ref{4.4}),
\begin{equation}\label{4.4b}
s=s_{anom}-s_{nc}\, .
\end{equation}
\begin{equation}\label{4.3b}
s_{nc}=\lim_{n\to 1}{nA_3^{nc}(1)-A_3^{nc}(n) \over n-1}~~,
\end{equation}
\begin{equation}\label{4.2b}
A_3^{nc}={1 \over 384 (4\pi)}\int_{\partial {\cal M}}\sqrt{H}d^2x~
\mbox{Tr}~\left[-12 \chi R\right]\, ,
\end{equation}
where conical singularities are allowed to contribute in (\ref{4.2b}).
With the help of (\ref{ie-3}) one finds
\begin{equation}\label{4.2cs}
A_3^{nc}(n)=nA_3^{nc}(1)-{1-n \over 32 \cos\alpha}
\sum_k\mbox{Tr}~\chi_k\, .
\end{equation}
This yields
\begin{equation}\label{4.3d}
s_{nc}(\alpha)={6a \over \cos\alpha}~~,
\end{equation}
which reduces to (\ref{4.3}) at $\alpha=0$.

Both $f(\alpha)$ and $s_{nc}(\alpha)$ are singular at $\alpha \to \pi/2$.
The case $\alpha=\pi/2$ is degenerate since the partition of the system onto two parts cannot be defined. Note that $f(\alpha)$ is monotonically decreasing function and $f(0)=0$.

\section{Comments on higher dimensions}\label{bc6}

Let us make some comments on what happens in higher dimensions.  We are interested in the case when the entangling surface $\Sigma$ intersects
the boundary $\partial M$ so that $P=\Sigma\bigcap {\partial M}$. We then expect that in any dimension $d$ there is a contribution to the logarithmic term in
entanglement entropy due to $P$. We shall denote this contribution as $s_0(P)$.  The concrete form of $s_0(P)$ depends on how many conformal invariants may contribute to this term
in dimension $d$. Let us list the possible contributions. First of all we notice that ${\rm dim} P=d-3$ so that $s_0(P)$ is given by integral over $P$ of a geometric quantity
of dimension $(d-3)$. This dimension is odd if $d$ is odd and even if $d$ is even. Being considered as a subset in the boundary $\partial{\cal M}$ the intersection $P$ is a co-dimension 2 surface and therefore there are 2  vectors
$p_a^\mu$, $a=1,2$ normal to $P$ in $\partial M$. The respective extrinsic curvature is $k^a_{\mu\nu}$, $a=1,2$. By projecting the extrinsic curvature $K_{\mu\nu}$ of the boundary $\partial M$ on these orthogonal vectors we construct $\hat{K}_{ab}=p_a^\mu p_b^\nu \hat{K}_{\mu\nu}$, where $\hat{K}_{\mu\nu}=K_{\mu\nu}-\frac{1}{d-1}H_{\mu\nu}K$ is trace-free extrinsic curvature. Then, the possible contributions to $s_0(P)$ are the bulk Weyl tensor both projected on normal vector $N^\mu$ and the normal vectors $p_a^\mu$,
$W_{anbn}$ and $W_{abcd}$, the conformal invariants constructed from the extrinsic curvature of the boundary, $\hat{K}_{\mu\nu}$, and its projections on vectors $p_a^\mu$, $\hat{K}_{ab}$, and also the trace free extrinsic curvature of $P$, $\hat{k}^a_{\mu\nu}=k^a_{\mu\nu}-\frac{1}{d-3}\gamma_{\mu\nu}k^a$, $\gamma_{\mu\nu}$ is the induced metric on 
$P$. We will not attempt here give a general form for the possible contributions to $s_0(P)$ since their number is rapidly growing with spacetime dimension.
The possible  terms in $s_0(P)$ in $d=4$ were suggested in \cite{Fursaev:2013mxa}.

\section{Concluding remarks}\label{bc7}

In this paper we have demonstrated that  there is a non-trivial logarithmic term in entanglement entropy
of a conformal field theory in $d=3$ dimensions provided the entangling surface intersects the boundary of the spacetime.
This feature is not typical only for three dimensions and it appears in any higher dimension $d$ when there is a non-trivial intersection $P$, see \cite{Fursaev:2013mxa}.
In odd dimensions, when the standard contribution is vanishing, there always exists a non-trivial logarithmic term due to $P$.

In dimension $d=3$ we have found that there is a certain relation between the logarithmic term in the entropy and the conformal charges in the integral trace anomaly.
However, this relation is less straightforward than in dimension $d=4$. We have demonstrated that the non-minimal coupling in the conformal field operator should be properly taken into account.

We have analyzed the conical geometry induced on the boundary by a conical singularity in the bulk. This geometry, both intrinsic and extrinsic,
appears to be rather non-trivial in the case when the entangling surface $\Sigma$ intersects the boundary under angle different from $\pi/2$.
These findings indicate that the conical geometry still has some surprises for us and perhaps more interesting features will be revealed in the future. 
 
Regarding the interplay of anomalies, entropy and the boundaries which is the main focus of the present paper we can mention at least two
interesting and important open questions:

\medskip

\noindent 1)  How the boundary charges which appear in the integrated Weyl anomaly
can be derived from the $n$-point correlation functions of
the CFT stress-energy tensor?

\medskip

\noindent 2) What is  the holographic description of the  boundary
terms in the Weyl anomaly and in the entanglement entropy? 

\medskip

The work towards the understanding the answer on the second question is on-going and we hope to report the progress
in the nearest future.

\bigskip

\noindent {\bf Acknowledgment}

\medskip

S.S. thanks Amin Astaneh and Clement Berthiere for fruitful discussions and the Dubna State University for warm hospitality
during the initial stage of the project.


\end{document}